**Transnational Network Dynamics of Problematic Information Diffusion**

Corresponding author:

Esteban Villa-Turek

Collaborators:

Rod Abhari

Erik C. Nisbet

Yu Xu

Ayse Deniz Lokmanoglu

**Abstract**

This study maps the spread of two cases of COVID-19 conspiracy theories and misinformation in Spanish and French in Latin American and French-speaking communities on Facebook, and thus contributes to understanding the dynamics, reach and consequences of emerging transnational misinformation networks. The findings show that co-sharing behavior of public Facebook groups created transnational networks by sharing videos of Médicos por la Verdad (MPV) conspiracy theories in Spanish and hydroxychloroquine-related misinformation sparked by microbiologist Didier Raoult (DR) in French, usually igniting the surge of locally led interest groups across the Global South. Using inferential methods, the study shows how these networks are enabled primarily by shared cultural and thematic attributes among Facebook groups, effectively creating very large, networked audiences. The study contributes to the understanding of how potentially harmful conspiracy theories and misinformation transcend national borders through non-English speaking online communities, further highlighting the overlooked role of transnationalism in global misinformation diffusion and the potentially disproportionate harm that it causes in vulnerable communities across the globe.

Keywords: social networks; health misinformation; global vulnerable communities; transnationalism; risk

**Transnational Network Dynamics of Problematic Information Diffusion**

Global viral diseases provide fertile ground for conspiracy theories, a prominent form of online misinformation, to spread (Cheng et al., 2021; Wood, 2018). For instance, COVID-19 conspiracy theories have been studied as a global phenomenon affecting predominantly English-speaking countries without explicit consideration of territorial variance (Cox & Halpin, n.d.; Kauk et al., 2021; Kearney et al., 2020; Yang et al., 2021). This is especially true regarding the transnational dimensions which contribute to the emergence and diffusion of conspiracy theories and misinformation at specific time points and geographic locations. This has led to a Westernized conceptualization of conspiracy theories and misinformation that assumes that the psychological dynamics and social conditions of the latter in Western, Educated, Industrialized, Rich, and Democratic (WEIRD) populations are generalizable to other populations (Henrich et al., 2010). It has been shown, however, that problematic and potentially harmful information has spread rapidly in online settings across non-WEIRD countries: in Latin America, for example, COVID-19 misinformation has been viral on social media, namely WhatsApp and Facebook (Valencia, 2021).

This study analyzes the online diffusion of COVID-19 conspiracy theories and misinformation across the Global South by comparing two cases of problematic information spread across public Facebook groups during the first 18 months of the pandemic and that vary in the cultural-linguistic dimension. The first case looks at the diffusion of COVID-19 conspiracy theories among Latin American countries and Hispanic communities on Facebook by studying the proliferation and diffusion of content related to the *Médicos por la Verdad*[1] (MPV) organization. The second case serves as a comparison and looks at a similar COVID-19 problematic information diffusion across francophone countries, in particular the one fueled by misleading remarks made early in the pandemic by French microbiologist Didier Raoult (DR) regarding the alleged efficacy of hydroxychloroquine to treat patients infected with the virus (Sayare, 2020).

---

[1] Spanish for *Doctors for Truth*

The importance of both of these cases relies on the fact that the online spread of problematic information in times of heightened uncertainty and possibility of risk increase is twofold. Not only is it comparatively understudied in non-WEIRD contexts, meaning that the Global South, representing a significant proportion of the world's population, has become a fertile ground for problematic information to blossom unchecked. It is also important to note that the peoples of the Global South are more vulnerable than those who live in WEIRD countries, meaning that the risk posed by their exposure to problematic online information is extremely disproportionate.

## Theoretical Framework

The notion of transnationalism in social media content diffusion is certainly not new. For instance, it has been employed to study the networks of far-right actors on Twitter to show how like-minded people across national borders form communities based on a mutually understood, shared worldviews and interpretative frames (Caiani & Della Porta, 2011; Froio & Ganesh, 2019). In other words, transnational online communication represents an additional channel for groups to consolidate narratives and audiences, which may in turn translate into offline activism and mobilization efforts, like those arranged and promoted by MPV in European and Latin American countries (Gardel, 2020; Maldita.es, 2021).

Increasingly, social networking sites (SNS) enable and mediate instances of human communication forming networks that are virtually unknown and in which flow is much harder to track, let alone to predict. This has brought about what has been termed the "dark social" (Madrigal, 2012; Swart et al., 2018) and we argue that we find ourselves in the age of "dark networks". Instances of this phenomenon can be found in the emergence of online communities of different sizes, on different platforms and for different purposes. In general, the theoretical emphasis around defining online communities has revolved around their two main components: what communities are and their online nature.

Regarding the first element, organizational theorists have argued that organizations precede communities, but the case has been made for postindustrial communities to precede and mediate the emergence of organizations (O'Mahony & Lakhani, 2011). Most importantly, however, is the idea that these contemporary forms of community can be better understood from a procedural standpoint by characterizing them as "a voluntary collection of actors whose interests overlap and whose actions are partially influenced by this perception" (O'Mahony & Lakhani, 2011). These communities therefore enable the convergence of a multitude of diverse backgrounds, interests, locations, etc., under the banner of some narrowly conceived perception of overlapping interests (O'Mahony & Lakhani, 2011), which mainly occurs due to the pervasiveness of networked digital communication technologies.

With regards to the second element, their digital nature, online communities can be thought of as those who exhibit the properties of pervasive awareness and persistent contact (Hampton, 2016). Both properties relate to the *person-to-network* (Hampton, 2016) nature of online communities, in which actors are perennially connected across time and space, and where actors can broadcast information without the ability to confirm its reception (Hampton, 2016).

Online communities therefore allow us to understand how problematic information was able to flow from the Global North to the Global South during the coronavirus pandemic. In other words, by allowing otherwise very dissimilar and distant actors to establish affinity ties by means of having overlapped and narrow interest (as signaled by being affiliated to public Facebook groups), online communities enabled a type of collective action - connective action - which was characterized by the existence of key groups that acted as transnational brokers. The theory of connective action (Bennett & Segerberg, 2012) was introduced in an effort to make sense of the multitude of digitally enabled organizational phenomena that were being sparked across the globe, from the *Arab Spring* to *Occupy Wall Street* to *los indignados*. More traditional theories on the emergence and organizational logic of offline mobilizations, such as the theories of collective action, were not able to fully explain the new geographically distant yet paradoxically unified instances of social mobilizations happening around very diverse issues across the world. In other words, the theory of connective action helped bridge the gaps that were starting to become evident with cases in which digital communication platforms afforded new kinds of networked communities, characterized by pervasive awareness and persistent contact (Hampton, 2016).

Thus, the theory of connective action enabled the characterization and theorization of particular communities that have two distinctive characteristic (Bennett & Segerberg, 2012). First, they are unified, not around a central organizing entity or idea, but around personalized and actionable ideas that are intended to relate to individuals at personal scale through the perception of narrow, overlapping interests (O'Mahony & Lakhani, 2011), even if there are other overarching differences among those individuals. This, in turn, eliminates some of the barriers that might otherwise inhibit the formation of communities (O'Mahony & Lakhani, 2011) in general and transnational online movements (Hampton, 2016) in particular. Second, the organizations that emerge following the logic of connective action should have a technological component such that it enables communication among those individuals that connect around shared personalized ideas (Bennett & Segerberg, 2012).

Although previous theories of collective action were not able to fully explain the existence of these technologically enabled cases of online communities, as (Bennett & Segerberg, 2012) propose, the logics of collective and connective action are not mutually exclusive but rather allow for a spectrum of hybrid categories of action logics. This hybridity is useful in cases where, although some elements of connective action logic are present, others are distinctively characteristic of the collective action logic, like the existence of concomitant propinquity or spatial proximity that could help connective action communities grow and thrive (O'Mahony & Lakhani, 2011) through the use of collective action elements like traditional resource mobilization (such as on-the-ground rallies). Another instance of hybridity can be observed in the fluidity of connective action roles (Bennett & Segerberg, 2012). As we will see, the seminal advocacy roles were fulfilled by more prominent and central actors, specifically those who ignited the spark by means of their perceived role of authority and expertise. These initial, somewhat static, roles did resemble those of traditional collective action theory, whereas the dynamic emergence of key transnational brokers resembled the connective action logic more.

The two case studies analyzed here involved problematic information, in the form of COVID-19 problematic information, that spread transnationally across countries and continents. This occurred specifically because they were enabled to do so by an underlying digital communication platform that allowed users to find and join groups of like-minded users around personalized and narrow interests (Bennett & Segerberg, 2012; O'Mahony & Lakhani, 2011), despite being located in distant geographies or having nothing else in common.

The possibility of networked yet diverse and possibly distant actors was possible therefore thanks to the connective affordances that digital communication platforms represent (Vaast et al., 2017). In such cases, the platform affords otherwise dissimilar individuals to connect and form networks based on shared interests or around distinctive issues. In these networks, roles are not fixed like in traditional theories of collective action, but rather allow for a fluid and dynamic emergence and cessation of roles in the network, like in the case of roles of support, advocacy and amplification (Vaast et al., 2017).

We argue that Facebook public groups are the quintessence of connective affordance: they are sets of very diverse actors that nevertheless coalesce around a very specific issue of interest, that in turn informed the creation of the group in the first place. In fact, studies have shown that, in general, Facebook users rely on the platform to connect with otherwise unknown people while at the comfort of home, in other words to expand their networks' weak ties in a form of bridging social capital (Papacharissi & Mendelson, 2010).

In fact, recent studies have found that Facebook groups afford several different kinds of connective action, for better or worse. For instance, a study of an extremist Swedish Facebook group found that the platform's architecture enabled the digitization of old anti-immigration discourse by allowing users to connect and assume different roles in an effort to increase the group's persuasive power (Merrill & Åkerlund, 2018). On the other hand, several other studies have documented how Facebook groups afford connective action for information seeking purposes. Specifically, another study of a Facebook group of international (non-Swedish) mothers living in Sweden found that, through the group's brokerage, otherwise dissimilar women were able to connect based on their shared motherhood in a foreign land to search and provide information relevant to their wellbeing (Mansour, 2021). Similarly, another study found that a Facebook group of Iranian women denouncing their experiences with hijab norms in the country afforded the creation of a crucial collective identity in the protest build-up process (Khazraee & Novak, 2018).

Therefore, it follows that Facebook groups can be understood as critical elements for connective action in that they broker information diffusion among otherwise geographically distant people who nevertheless overlap around narrow and shared particular interests represented by the thematic nature of the Facebook groups. By employing a network perspective, we are able to identify dynamic brokerage roles similar to those of connective action (advocacy, support or amplification), but with the crucial function of operating at a transnational level.

We argue that distinctively active groups served as transnational brokers in the process of problematic information diffusion between the Global North and the Global South during the coronavirus pandemic. This is the case, because the fluid nature of emergent roles in the logic of connective action allows for actors (public Facebook groups in this study) to broker between otherwise distant countries and continents. In fact, brokerage roles have been found to play a critical role in social movements by fulfilling the need to bridge between otherwise disconnected actors, therefore ensuring cohesion among the collective's members (Diani & McAdam, 2003). This potential brokerage role, however, has been found to depend greatly on certain actor attributes (Diani & McAdam, 2003). In fact, we expect influential groups to operate enabled by the fact that there is a shared linguistic element (a "postcolonial affordance" of sorts), which supported the emergence of the transnational misinfodemic network in the first place.

Therefore, we set out to investigate which group characteristics explain why they became or not influential actors in the diffusion process and whether the same group-level characteristics are able to explain the emergence of ties in the network as a whole. As has been explained above, online communities operate by linking dissimilar actors under a narrow and personalized overlapping area of interest (O'Mahony & Lakhani, 2011). Thus, we hypothesize first that thematic similarity between groups increases the likelihood of tie formation in the network, since the initial narrow yet shared issue of interest informed the creation and growth of the groups in the first place.

Second, it has been noted that the logic of connective action allows for hybridity in connective affordances, meaning that online communities can also be strengthened by physical proximity (Bennett & Segerberg, 2012; O'Mahony & Lakhani, 2011). To capture possible geographical proximity effects in the emergence of the network, we employ a measure of geographic location of the groups at the country level, hypothesizing that geographic proximity among groups increases the likelihood of tie formation in the network.

Third, we argue that there are latent cultural similarity effects among groups, and that they increase the likelihood of tie formation in the network. For this purpose we employ a measure of that captures cultural similarities among countries well, Facebook's Social Connectedness Index (SCI) (Bailey et al., 2018). In fact, the SCI has proven useful in correlating transnational connectedness with historic intranational and international migratory patterns (Bailey et al., 2018), as well as international trade patterns (Bailey et al., 2021), capturing therefore intangible relationships that serve as a proxy for cultural proximity in this study.

Finally, as noted above, the appearance of influential roles in the network, such as brokerage, is conditional on the existence of certain characteristics in the actors (Diani & McAdam, 2003). Therefore, given that shared linguistic attributes among postcolonial countries is a sine qua non characteristic for the emergence of the networks under study, it is necessary to include them as group-level attributes to estimate their role in its emergence, arguing that a shared language increases the likelihood of tie formation and, subsequently, enables the emergence of dynamic roles within the network.

This study uses a network approach that is sensitive to transnational communication dimensions, specifically the geographic proximity between Facebook groups, the cultural proximity of audiences in diverse parts of the world, all while upholding the importance of shared linguistic attributes in the process. The latter is a crucial element for understanding the rapid diffusion of COVID-19 problematic information in Spanish and French, which emerged in Europe and spread rapidly across Latin American and francophone countries, respectively. Isolating the network structure of conspiracy theory and misinformation diffusion exposes their distinctive structural characteristics, which enables researchers, healthcare workers and policymakers to develop targeted responses and contributes to a better evidence-based approach to global communication governance efforts (GCG) (Padovani & Pavan, 2011).

## Methodology

### *Data, sampling and measures*

The units of analysis are public Facebook groups that shared native or externally hosted audiovisual content that included mentions to the key terms *Médicos por la Verdad* or *Natalia Prego* (one of the group's founders and most vocal member) for the MPV case in Spanish language, and *Didier Raoult* or *hydroxychloroquine*, for the DR case in French language. The data were queried using CrowdTangle, a tool developed by Facebook that allows researchers to obtain and analyze public data shared on the social media platform (Bleakley, 2021). Our case studies center therefore on the Facebook network of groups that shared videos featuring problematic information across the Spanish and French speaking postcolonial world. Combined, the Facebook groups amassed audiences exceeding several orders of magnitude.

*Case 1: Médicos por la Verdad and the false pandemic*

The first case looks at conspiracy theories in Spanish that were promoted by the organization *Médicos por la Verdad* (MPV), based in Spain and accruing significant reception across Latin American countries throughout the pandemic (Maldita.es, 2021). Mentions of MPV reached close to 16,000 across public Facebook groups from Spain and Latin America during the pandemic, and in 2020 alone local chapters of the organization began emerging in more than 10 countries (Gardel, 2020). Due to their viral audiovisual diffusion on social media (Knuutila et al., 2020), MPV conspiracies have posed a significant increase in public health risks in Latin America and other Hispanic diasporas across the continent by promoting the use of unauthorized medicine, challenging the efficacy of face masks, doubting the accuracy of testing kits, all while arguing that the pandemic is a "false pandemic" or a "plandemic".

The study analyzes Facebook data of non-geolocated posts made by public groups or pages containing mentions of a prominent COVID-19 conspiracy organization from Spain and of one of its most vocal members. The Facebook data is non-geolocated to allow for posts that lack this metadatum to also be included in the query results. The complete dataset contains posts from February 1, 2020, until October 1, 2021, with video content (native, native live, and externally hosted videos) for a total of around 15,336 posts. These data are used to build a bipartite network where nodes are Facebook groups and video URLs, and a directed link is formed between them when a group or page shares a URL. The full network comprises 5781 unique nodes and 2243 unique video URLs. Based on the number of times a node shares a video, a cutoff threshold is set to retain only those groups that participated in video co-sharing behavior at least twice during the sampling period to ensure minimal sharing behavior in the subsequent analysis. This operation yields a directed bipartite network composed of 2168 nodes and 1968 unique video URLs.

This bipartite network is then projected onto a one-mode co-share undirected network where Facebook groups are nodes and ties are formed between two nodes if they have co-shared the same video URL. To focus only on the super-spreader core of the network, we trimmed it based on edge weight (i.e., the times that two groups co-shared a video), retaining only the edges whose weight belongs to the upper percentile of the edge weight distribution. The resulting undirected one-mode network contains 518 nodes and 2390 edges.

The resulting super-spreader network was then manually labeled to classify each node according to their geographical location, the language of their posts and their category as given by the overarching theme of the group's activity and content. Due to the limited availability of metadata about several of the Facebook groups, a series of indicators were used to assign the labels. For the *language* label, indicators in the group's name, description or recent activity were used to determine the most common (if not unique) language used and the corresponding language ISO codes were used as labels. The country or *domain* label was assigned by looking at indicators like the mention of a geographical location (country, city, town, region, etc.) in the group's name or description, or, alternatively, by the nationality of the group's administrators, depending on availability, after which the country ISO codes were employed as labels. Finally, the *group category* label was assigned by reviewing a mixture of indicators whenever they were found in the group's name, the group's description and/or most recent group's posts and activity, depending on their availability for each group. Then, each group was assigned to five possible categories: Conspiracy Theories, Media, Politics, Religion & Spirituality, and Community/Other. After removing those groups whose country was not included in the SCI (such as Russia, Cuba or Venezuela) and those that could not be reasonably inferred, we retained 467 nodes and 2170 edges. Figures 1, 2 and 3 showcase the distribution of nodal attributes in the MPV network data.

**Figure 1**

*Distribution of countries in MPV data*

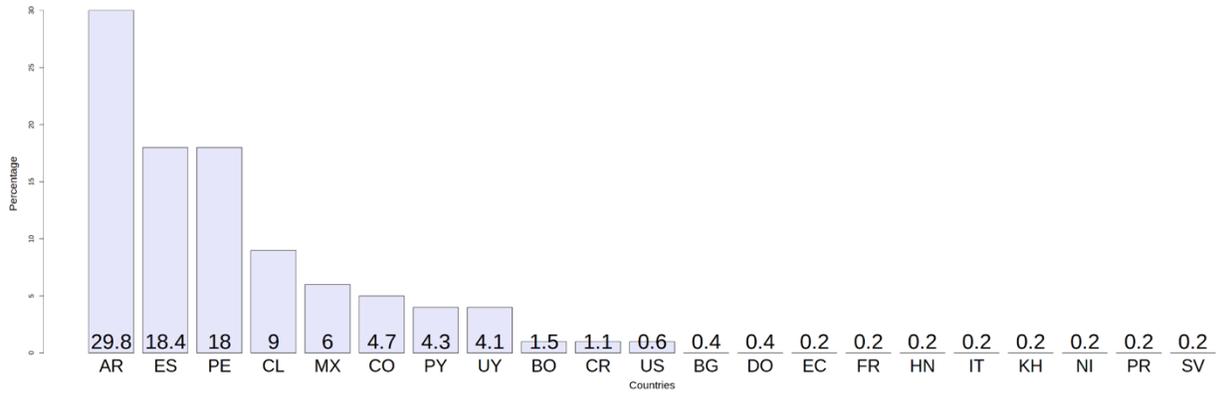

**Figure 2**

*Distribution of languages in MPV data*

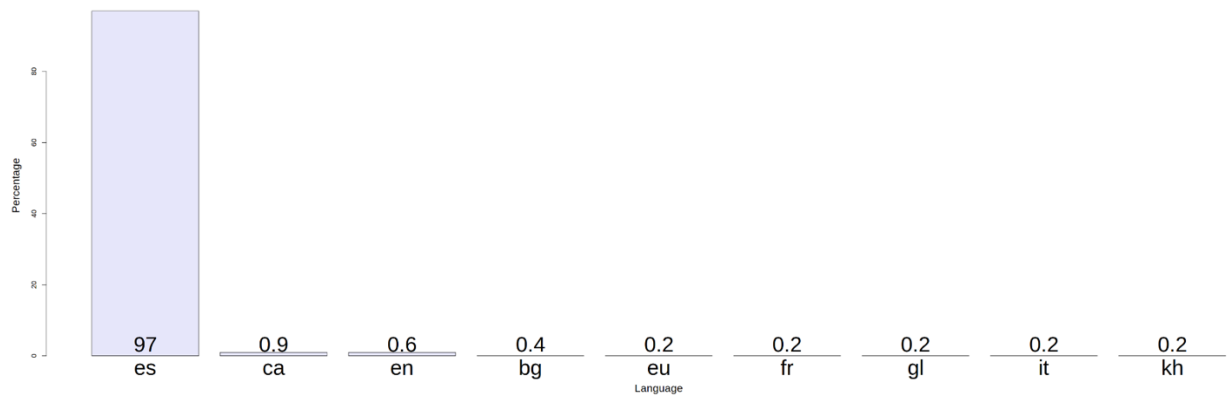

**Figure 3**

*Distribution of group themes in MPV data*

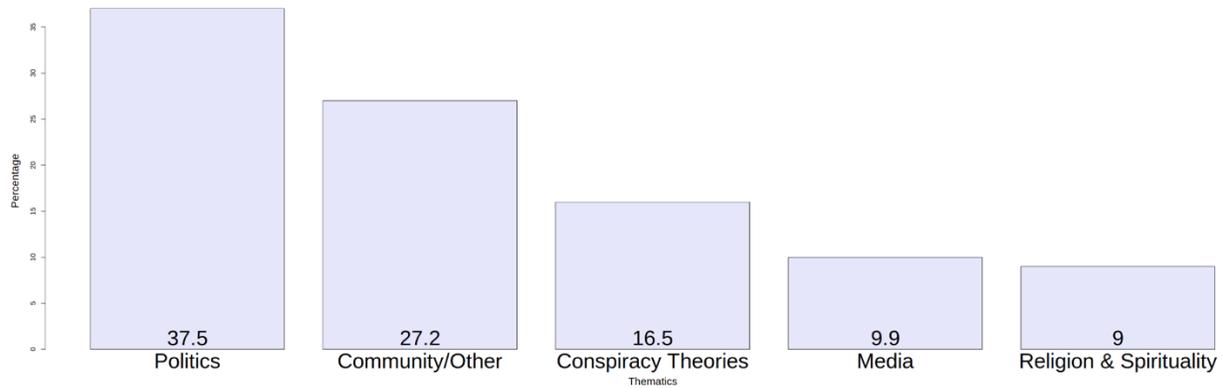

*Case 2: Didier Raoult and the alleged effectiveness of hydroxychloroquine*

The second case looks at problematic information in French that was ignited by early declarations made public by the controversial French microbiologist Didier Raoult, who promoted the use of hydroxychloroquine to effectively treat patients infected with the coronavirus (Sayare, 2020). The data analyzed for this case comprises Facebook data of non-geolocated posts made by public groups containing mentions of "Didier Raoult" and "Hydroxychloroquine" from February 1, 2020, until October 1, 2021, with audiovisual content (native, native live, and externally hosted videos) for a total of 11,122 posts.

Following the exact same procedure as with the MPV case data, we built a bipartite network where nodes were Facebook groups and video URLs, and a directed link is formed between them when a group or page shares a URL. The full network comprised 1170 unique nodes and 1195 unique video URLs. This bipartite network was also projected onto a one-mode co-share undirected network where groups and pages are nodes and ties are formed between two nodes if they have co-shared the same video URL. The largest component of the resulting undirected one-mode network contained 1156 nodes and 120274 edges. As in the MPV case, edges were trimmed based on their weight to retain only the upper percentile of heaviest edges, that is, those that account for most of the co-sharing behavior in the network. After removing nodes whose country could not reasonably be inferred, the resulting subset of the network comprised 454 nodes and 1191 edges.

This subset was also manually labeled to classify each node according to their geographical location, the language of their posts and their category as given by the theme of their activity. Due to the same limitations in available information in several of the Facebook groups, a series of heuristics were used to assign the labels. For the *language* label, cues in the group's name, description or recent activity were used to determine the most common (if not unique) language used and the corresponding language ISO codes were used as label. The country or *domain* label was assigned by looking at cues like the mention of a geographical location (country, city, town, region, etc.) in the group's name or description, or, alternatively, by the nationality of the group's administrators, depending on availability, after which the country ISO codes were employed as labels. Finally, the *group category* label was assigned by reviewing a mixture of cues available in the group's name, description or most recent activity and assigned to five possible categories: Conspiracy Theories, Media, Politics, and Community/Other (unlike the MPV data, there were no groups with overarching Religion & Spirituality themes in the DR data). Figures 4, 5 and 6 showcase the distribution of nodal attributes in the DR network data.

**Figure 1**

*Distribution of countries in DR data*

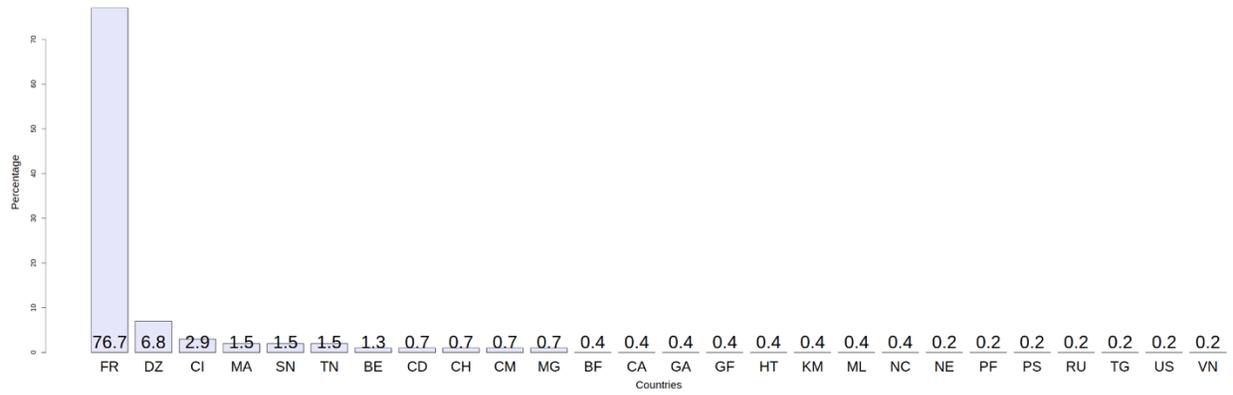

**Figure 2**

*Distribution of languages in DR data*

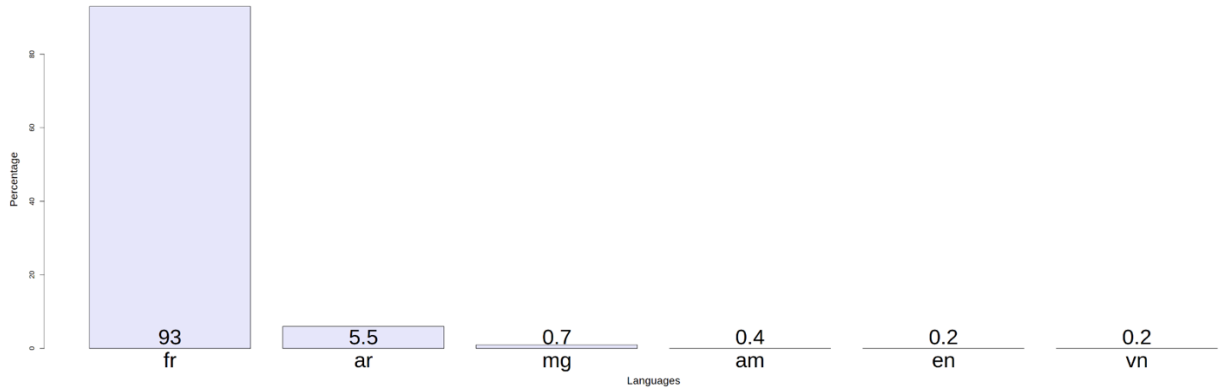

**Figure 3**

*Distribution of group themes in DR data*

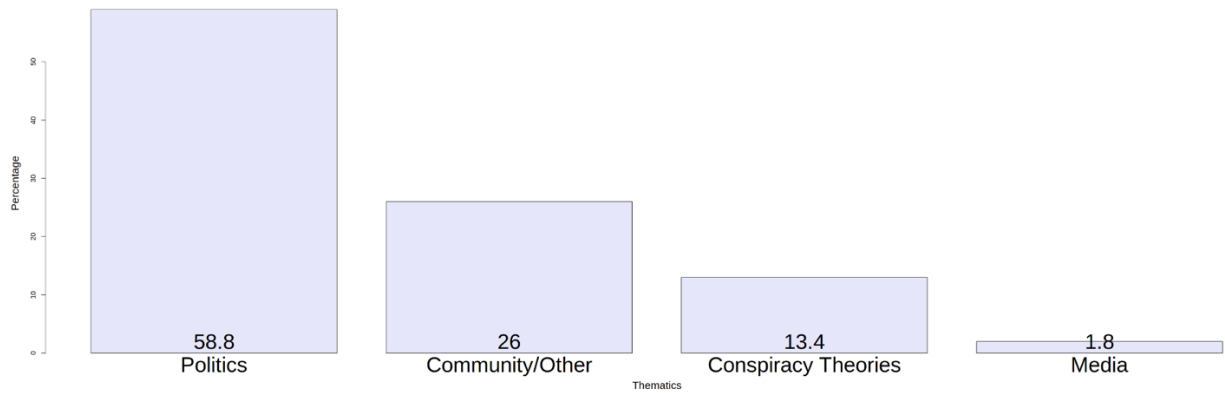

## Statistical Analysis and Results

After analyzing the main brokers in the network qualitatively, the study employed Exponential Random Graph Models (ERGMs) to statistically determine which nodal attributes are significant in explaining the emergence of network ties. ERGMs are particularly well suited for this task, as they allow researchers to statistically assess and predict the structural characteristics of ties in an observed network. This operates by simulating randomly-generated networks that are specified to structurally equate the observed network, allowing to conduct hypothesis tests on the observed network's attributes (Robins & Lusher, 2013). In other words, ERGMs are useful to compare in this study the relative influence of transnational factors on tie formation in the networks to other non-transnational actor-level factors to contrast their predictive power around the emergence dynamics of the networks (Robins et al., 2007). In fact, ERGMs have been emphasized in the literature as an ideal method for social network analysis in communication studies, mainly because they allow researchers to employ structure, external relationships or attributes as predictors, all while accounting for the interdependent nature of network data (Shumate & Palazzolo, 2010). Therefore, the study estimates two ERGMs on the two networks using the *ergm* package version 3.9.4, as implemented for the R programing language (Handcock et al., 2018).

To estimate ERGMs for both networks, the study makes use of endogenous attributes as mechanisms through which the network is structured, as well as of exogenous attributes of the nodes and of exogenous relations in the network (Contractor et al., 2006). In general, endogenous attributes correspond to those self-organizing factors in a network that are merely structural and do not depend on attributes of nodes or edges, like the number of edges or triangles. Conversely, exogenous attributes are the ones that help explain the formation of ties in a network based on non-structural attributes, like attributes of nodes and edges, or other relationships between the nodes that exist in other networks, and that could help explain the emergence of ties in the observed network. We hypothesize, as mentioned above, that linguistic similarity, thematic similarity, and cultural and geographic proximity increase the likelihood of tie formation in both networks. If the hypotheses hold, it would be expected to see a higher likelihood in co-sharing activity in the network between groups that share the same language, the same themes, and that are culturally or geographically close to one another, as captured by the SCI (Bailey et al., 2018).

Exogenous attributes are employed to estimate the influence of non-structural characteristics of the network on its emergence, like node characteristics or external network relationships. The exogenous nodal attributes to test for linguistic and thematic similarity are given by the hand-coded *language* and *group category* labels introduced above. The exogenous relations in the network to account for cultural similarity are modeled using the SCI, which is calculated by taking the number of Facebook connections between two geographic regions and dividing them by the product of Facebook users in both regions, which is then scaled to have a maximum value of 1,000,000,000 (maximum connectedness) and 1 (minimum connectedness), accounting for every possible country-country pair and without having any missing values (*Facebook Data For Good Social Connectedness Index Methodology*, 2021). This study employs the SCI to account for cultural similarity relations in social networks, especially those that emerge across national borders. As mentioned above, the SCI empirically shows that people across the world tend to form ties with those who are most similar to them, based for instance on sociodemographic characteristics like race, age, religion, education, but also on other "intangible aspects such as attitudes and beliefs (…)", which can be understood, among others, as cultural in nature (Bailey et al., 2018). To reiterate, the SCI has proven useful to correlate transnational connectedness with historic intranational and international migratory patterns (Bailey et al., 2018), a relationship that might help account for the emergence of ties between seemingly distant (or close) countries in both networks under study. Finally, the nodal attribute to model geographic co-location is given by the hand-coded *domain* label discussed above.

*MPV ERGM*

Table 1 presents the estimated ERGM coefficients.

**Table 1**

*ERGM coefficients for the MPV network*

|  | Coefficient |
|---|---|
| edges | −3.483*** |
|  | (0.120) |
| edgecov.sciNet_statnet | 0.009* |
|  | (0.003) |
| nodematch.domain | 0.063 |
|  | (0.058) |
| nodematch.language | −1.573*** |
|  | (0.057) |
| mix.groupCat.Community/Other.Conspiracy Theories | 1.050*** |
|  | (0.132) |

|  | Coefficient |
|---|---|
| mix.groupCat.Conspiracy Theories.Conspiracy Theories | 1.719*** |
|  | (0.145) |
| mix.groupCat.Community/Other.Media | 0.275 |
|  | (0.171) |
| mix.groupCat.Conspiracy Theories.Media | 1.169*** |
|  | (0.157) |
| mix.groupCat.Media.Media | 0.599* |
|  | (0.293) |
| mix.groupCat.Community/Other.Politics | 0.569*** |
|  | (0.128) |
| mix.groupCat.Conspiracy Theories.Politics | 1.594*** |
|  | (0.123) |
| mix.groupCat.Media.Politics | 0.968*** |
|  | (0.142) |
| mix.groupCat.Politics.Politics | 1.250*** |
|  | (0.127) |

|  | Coefficient |
|---|---|
| mix.groupCat.Community/Other.Religion & Spirituality | 0.127 |
|  | (0.183) |
| mix.groupCat.Conspiracy Theories.Religion & Spirituality | 1.061*** |
|  | (0.164) |
| mix.groupCat.Media.Religion & Spirituality | −0.044 |
|  | (0.292) |
| mix.groupCat.Politics.Religion & Spirituality | 0.789*** |
|  | (0.149) |
| Num.Obs. | 108811 |
| AIC | 20244.6 |
| BIC | 20407.7 |

The interpretation of ERGM coefficients is similar to that of a logistic regression. In that sense, they allow us to assess whether each term increases or decreases the likelihood of tie formation in the networks.

In terms of geographical co-location, the results do not indicate a statistically significant effect on the likelihood of tie formation. However, the same cannot be said of cultural similarity, which shows a positive and significant effect, indicating that cultural similarity between groups' nationalities increases the likelihood of tie formation in the network, holding everything else constant. Regarding language similarity, our findings indicate that linguistic similarity decreases the likelihood of tie formation, holding everything else constant. In terms of thematic similarity, we estimated the likelihood of tie formation among all possible thematic pairs in the network. The results range from the intuitive increase in likelihood of tie formation in pairs like Conspiracy Theory-Conspiracy Theory groups, or Politics-Politics groups. Nevertheless, our estimates indicate that other possible thematic pairs that include Conspiracy Theory groups see an increase in likelihood of tie formation, such as in the cases of Conspiracy Theory groups paired with Community, Media, Politics or Religion & Spirituality groups, which seems to indicate a sort of spillover effect on the diffusion of problematic information from dedicated Facebook groups to non-dedicated conspiratorial groups.

**DR ERGM**

Table 2 presents the estimated ERGM coefficients.

**Table 2**

*ERGM coefficients for the DR network*

|  | **Coefficients** |
|---|---|
| edges | −9.232*** |
|  | (1.002) |
| edgecov.sciNet_statnet | 1.672*** |

|  | Coefficients |
|---|---|
|  | (0.059) |
| nodematch.domain | 0.588*** |
|  | (0.084) |
| nodematch.language | −0.185 |
|  | (0.129) |
| mix.groupCat.Community/Other.Conspiracy Theories | 4.381*** |
|  | (1.004) |
| mix.groupCat.Conspiracy Theories.Conspiracy Theories | 4.947*** |
|  | (1.008) |
| mix.groupCat.Community/Other.Media | 2.535* |
|  | (1.225) |
| mix.groupCat.Conspiracy Theories.Media | 4.504*** |
|  | (1.055) |
| mix.groupCat.Community/Other.Politics | 3.465*** |
|  | (1.001) |
| mix.groupCat.Conspiracy Theories.Politics | 4.459*** |

|  | Coefficients |
|---|---|
|  | (1.001) |
| mix.groupCat.Media.Politics | 3.539*** |
|  | (1.032) |
| mix.groupCat.Politics.Politics | 4.039*** |
|  | (1.000) |
| Num.Obs. | 102831 |
| AIC | 11744.2 |
| BIC | 11858.7 |

In the DR network data geographical co-location indicates a positive and statistically significant effect on the likelihood of tie formation, as well as cultural similarity, which shows a positive and significant effect. In this regard the estimates differ from the MPV network, which could be attributed to the higher population variance in terms of culture, language, geography, etc. In the same vein, it seems that language similarity, would increase the likelihood of tie formation in the network, but our estimate is not statistically significant. Finally, in terms of thematic similarity, the same trend operates on all possible thematic pairs in the network. In that sense, we see the same intuitive increase in likelihood effect in Conspiracy Theory-Conspiracy Theory, and Politics-Politics groups. More interestingly, our estimates for this network also indicate that other thematic pairs in which Conspiracy Theory groups are active see an increase in likelihood of tie formation. In other words, thematic pairs in which Conspiracy Theory groups are paired with Community, Media or Politics groups also see an increase in the likelihood of tie formation in the network (there were no Religion & Spirituality groups in the super-spreader core for the DR network). This finding, reiterating, would indicate a spillover effect on the diffusion of problematic information from dedicated groups to unassuming and vulnerable groups.

## Discussion and Limitations

This study compared two similar cases of online transnational diffusion of problematic information through Facebook groups using social network analysis techniques. Our findings suggest that cultural similarity among groups' nationalities has a noticeable effect on the likelihood of tie formation in both networks, whereas taking language similarity as predictor of the likelihood of co-sharing behavior yields a negative effect in the MPV data and a non-significant estimate in the DR network. Geographic proximity also varies between case studies, with likelihood of tie formation increasing with geographical co-location in the DR network but yielding non-significant estimates in the MPV network. These initial findings make intuitive sense. In the MPV case, we studied how problematic information in Spanish was manufactured in Spain and consumed throughout the pan-Hispanic world. As such, Spanish is spoken in all countries (which is why linguistic similarity yields a tendency against tie formation, holding everything else constant) and culturally speaking all countries are homogeneous to a large degree (given, for instance, the way in which Spanish colonization took place historically). The DR case study is a little different in this sense. Here, all countries are more culturally heterogenous, meaning that geographical co-location needs to be present in order to allow the flow of information among group nationalities. However in this case cultural similarity also indicates a tendency towards tie formation, again a possible remnant of another, yet historically different, colonization process.

Regarding our findings about the influence thematic similarity among possible thematic pairs, we can see the exact same tendencies towards tie formation in both the MPV and DR networks. This could be indicative of the superlative diffusion power that dedicated conspiratorial and misinforming Facebook groups have over other non-dedicated, unassuming groups and their audiences. A clear example of this phenomenon exists in the tendency towards tie formation between Conspiracy Theory groups and Community groups in both networks. The relevancy of the example lies in the fact that most Community groups encompass regular user who connect based on common hobbies, interests, or location, and therefore create online communities to share mutually beneficial information. When those groups co-share problematic information with dedicated misinformation-spreading groups, all unassuming participants in the Community groups are exposed to potentially harmful content without consenting to it.

A plausible limitation of the study is the focus on the networks' super-spreader cores since this approach excludes the possibility of examining core-periphery dynamics that could be present in both cases. Another substantial limitation is the generalizability of the findings. Although thematic similarity seems to be a common predictor of co-sharing activity in both networks, it remains to be seen if the same mechanism operates in other cases of transnational diffusion of problematic information. This study contributes to the increasingly relevant research expansion on online diffusion of problematic information beyond the usual WEIRD contexts, in order to venture into the much more diverse realm of the Global South. This is a critical step towards a better understanding of how potentially massive, networked audiences are exposed to problematic information that could pose risks of disproportionate harm to vulnerable communities around the world.

## Conclusions and Policy Implications

These findings highlight the complex interplay between linguistic, thematic, and cultural factors in shaping online problematic information diffusion. They underscore the need for targeted network-informed interventions to address its spread from the Global North to the Global South. By identifying key factors influencing tie formation, platform moderators and product policymakers can implement targeted interventions to mitigate the spread of extremist content. Overall, the study offers valuable insights and methodologies that can help online platforms develop more effective strategies for preventing the proliferation of online misinformation and extremism. By leveraging network analysis techniques and accounting for geographic, cultural, linguistic, and thematic similarities, platforms can enhance their ability to detect and mitigate extremist content, ultimately creating safer and more inclusive online environments.

*Identification of High-Risk Nodes*

Platforms can prioritize monitoring and intervention strategies by analyzing the structure of online networks and identifying high-risk nodes, such as Conspiracy Theory groups, which often serve as brokers in disseminating problematic information, including extremist content. By targeting these brokering agents within the networks, platforms can effectively disrupt the spread of extremism.

*Understanding Information Flow*

The study's network analysis techniques allow a deeper understanding of how information flows within online ecosystems. Platforms can leverage this knowledge to track the propagation of extremist narratives and identify key pathways through which they spread. By mapping out these pathways, platforms can implement targeted interventions to prevent the rapid dissemination of extremist content.

*Assessing Cultural and Linguistic Factors*

The study highlights the importance of cultural and linguistic factors in shaping online interactions for information sharing. Platforms can use this insight to tailor their moderation efforts to specific linguistic and cultural contexts. By understanding the unique dynamics of different communities, platforms can develop more effective strategies for combating extremism and misinformation.

*Monitoring Thematic Similarity*

The emphasis on thematic similarity as a significant factor in forming ties in online networks underscores the importance of monitoring specific topics and themes associated with misinformation and extremism. By tracking the spread of extremist narratives across thematic boundaries, platforms can detect emerging trends and preemptively intervene to prevent the escalation of extremist activity.

*Integrating Geographical Proximity*

In the case study focusing on French-speaking groups, geographical co-location emerged as a significant factor in tie formation. Platforms can incorporate this insight into their moderation efforts by considering the geographical distribution of users and the potential impact of local context on misinformation and extremist activity. By accounting for geographical proximity, platforms can develop more nuanced strategies for addressing extremism at the regional level.

*Applying Social Network Analysis Techniques*

Finally, the use of computational and statistical methods demonstrates the value of social network analysis techniques in understanding and predicting online behaviors. Platforms can leverage these techniques to identify patterns of extremist activity, predict future trends, and assess the effectiveness of intervention strategies. By adopting a data-driven and network-informed approach, platforms can better allocate resources and prioritize efforts to combat extremism.

TRANSNATIONAL NETWORK DYNAMICS OF DIFFUSION                                    31**References**

Bailey, M., Cao, R., Kuchler, T., Stroebel, J., & Wong, A. (2018). Social connectedness: Measurement, determinants, and effects. *Journal of Economic Perspectives*, *32*(3), 259–280. https://doi.org/10.1257/jep.32.3.259

Bailey, M., Gupta, A., Hillenbrand, S., Kuchler, T., Richmond, R., & Stroebel, J. (2021). International trade and social connectedness. *Journal of International Economics*, *129*, 103418. https://doi.org/10.1016/j.jinteco.2020.103418

Bennett, W. L., & Segerberg, A. (2012). The Logic of Connective Action. *Information, Communication & Society*, *15*(5), 739–768. https://doi.org/10.1080/1369118X.2012.670661

Bleakley, W. (2021). *About Us | CrowdTangle Help Center*. https://help.crowdtangle.com/en/articles/4201940-about-us

Caiani, M., & Della Porta, D. (2011). *The elitist populism of the extreme right: A frame analysis of extreme right-wing discourses in Italy and Germany w*. *46*, 180–202. https://doi.org/10.1057/ap.2010.28

Cheng, M., Yin, C., Nazarian, S., & Bogdan, P. (2021). Deciphering the laws of social network-transcendent COVID-19 misinformation dynamics and implications for combating misinformation phenomena. *Scientific Reports*, *11*(1), 1–14. https://doi.org/10.1038/s41598-021-89202-7

Contractor, N. S., Wasserman, S., & Faust, K. (2006). Testing multitheoretical, multilevel hypotheses about organizational networks: An analytic framework and empirical example.

TRANSNATIONAL NETWORK DYNAMICS OF DIFFUSION                                33Henrich, J., Heine, S. J., & Norenzayan, A. (2010). The weirdest people in the world? *Behavioral and Brain Sciences*, *33*(2–3), 61–83. https://doi.org/10.1017/S0140525X0999152X

Kauk, J., Kreysa, H., & Schweinberger, S. R. (2021). Understanding and countering the spread of conspiracy theories in social networks: Evidence from epidemiological models of Twitter data. *PLOS ONE*, *16*(8), e0256179. https://doi.org/10.1371/JOURNAL.PONE.0256179

Kearney, M. D., Chiang, S. C., & Massey, P. M. (2020). The Twitter origins and evolution of the COVID-19 "plandemic" conspiracy theory. *Harvard Kennedy School Misinformation Review*, *1*(3), 1. https://doi.org/10.37016/mr-2020-42

Khazraee, E., & Novak, A. N. (2018). Digitally Mediated Protest: Social Media Affordances for Collective Identity Construction. *Social Media + Society*, *4*(1), 205630511876574. https://doi.org/10.1177/2056305118765740

Knuutila, A., Herasimenka, A., Au, H., Bright, J., & Howard, P. N. (2020). Covid-Related Misinformation On Youtube The Spread of Misinformation Videos on Social Media and the Effectiveness of Platform Policies. *COMPROP Data Memo*, *6*, 1–7.

Madrigal, A. C. (2012, October 12). *Dark Social: We Have the Whole History of the Web Wrong*. The Atlantic. https://www.theatlantic.com/technology/archive/2012/10/dark-social-we-have-the-whole-history-of-the-web-wrong/263523/

Maldita.es. (2021). *The International Scheme of "Doctors for the Truth": A Denialist Trademark Registered by Natalia Prego—Chequeado*. Chequeado. https://chequeado.com/investigaciones/the-international-scheme-of-doctors-for-the-truth-a-denialist-trademark-registered-by-natalia-prego/

TRANSNATIONAL NETWORK DYNAMICS OF DIFFUSION                                    34

TRANSNATIONAL NETWORK DYNAMICS OF DIFFUSION 35Shumate, M., & Palazzolo, E. T. (2010). Exponential Random Graph (p*) Models as a Method for Social Network Analysis in Communication Research. *Communication Methods and Measures*, *4*(4), 341–371. https://doi.org/10.1080/19312458.2010.527869

Swart, J., Peters, C., & Broersma, M. (2018). Shedding light on the dark social: The connective role of news and journalism in social media communities. *New Media & Society*, *20*(11), 4329–4345. https://doi.org/10.1177/1461444818772063

Vaast, E., Safadi, H., Lapointe, L., & Negoita, B. (2017). Social Media Affordances for Connective Action: An Examination of Microblogging Use During the Gulf of Mexico Oil Spill. *MIS Quarterly*, *41*(4), 1179–1206.

Wood, M. J. (2018). Propagating and Debunking Conspiracy Theories on Twitter during the 2015-2016 Zika Virus Outbreak. *Cyberpsychology, Behavior, and Social Networking*, *21*(8), 485–490. https://doi.org/10.1089/cyber.2017.0669

Yang, A., Shin, J., Zhou, A., Huang-Isherwood, K. M., Lee, E., Dong, C., Kim, H. M., Zhang, Y., Sun, J., Li, Y., Nan, Y., Zhen, L., & Liu, W. (2021). The battleground of COVID-19 vaccine misinformation on Facebook: Fact checkers vs. Misinformation spreaders. *Harvard Kennedy School Misinformation Review*. https://doi.org/10.37016/mr-2020-78